# Origin of persistent photoconductivity in surface conducting hydrogenated diamond films

N. Mohasin Sulthana [a,b], K. Ganesan [a,b,1], P.K. Ajikumar [b]

[a] *Homi Bhabha National Institute, Training School Complex, Anushakti nagar, Mumbai-400094, India*

[b] *Materials Science Group, Indira Gandhi Centre for Atomic Research, Kalpakkam- 603102, India*

**Abstract**

The p-type surface conductivity of hydrogen-terminated diamond (HD) has opened up new possibilities for the development of diamond-based electronic devices. However, the origin of the persistent photoconductivity (PPC) observed in surface-conducting HD remains unclear—an understanding that is crucial for advancing HD-based optoelectronic technologies. In this study, we investigate the underlying mechanism of PPC in surface-conducting HD films. A systematic analysis was performed by tuning the carrier density via partial oxygen termination using an ozonation process. With increasing O-termination, both the decay time and the recombination barrier of photoexcited electron-hole pairs were found to decrease significantly—from 232 to 5 seconds, and from ~ 150 to 54 meV, respectively. Temperature-dependent measurements reveal that PPC in HD is influenced by random local potential fluctuations, which delay the recombination of photoexcited carriers. Furthermore, the observed PPC behavior is closely associated with percolative transport processes within the HD film. Importantly, the dependence of PPC on sheet carrier density is correlated with Coulomb interactions between the two-dimensional hole gas and the surface adsorbate layer. This study offers new insights into the PPC mechanism in surface-conducting HD films, contributing to the broader understanding necessary for the design of advanced diamond-based optoelectronic devices.

---

[1] Corresponding author. Email : kganesan@igcar.gov.in ( K. Ganesan)

# 1. Introduction

Diamond is considered one of the most promising wide bandgap semiconductors due to its exceptional properties, including an ultra-wide bandgap, high carrier mobility, excellent electric breakdown strength, superior thermal conductivity, biocompatibility, chemical inertness, and a wide potential window for electrochemical applications [1,2]. Despite these advantages, the development of diamond-based electronics faces key challenges, primarily its extremely high intrinsic resistivity and the limited availability of shallow dopants. To overcome these limitations, hydrogen termination of the diamond surface (H-diamond or HD) offers an effective approach to modulate surface conductivity (SC) via a charge transfer doping mechanism. This mechanism involves the interaction between the H-diamond surface and adsorbed species, resulting in the formation of a two-dimensional hole gas (2DHG) [3,4]. The presence of this 2DHG makes diamond particularly suitable for a range of advanced applications, including field-effect transistors, high-power and high-frequency devices, thermionic electron emitters, photodetectors, and chemical and biological sensors [2,5,6].

The performance of photodetectors is often constrained by sub-bandgap photoresponse and persistent photoconductivity (PPC)—a phenomenon in which conductivity persists even after the photoexcitation source is removed. In recent years, PPC has gained widespread attention for its applicability across various technological domains, serving as a key mechanism in optical memory devices, optoelectronic devices, solar energy conversion, artificial vision systems, and bioelectronic applications [7]. Gaining insight into the origin of PPC is essential for understanding the underlying charge storage mechanisms and relaxation processes. There are a few reports on PPC in diamond, each interpreted based on different underlying mechanisms [8–13]. In polycrystalline diamond, PPC is commonly associated with structural imperfections, including point defects, grain boundaries, and non-diamond carbon phases that introduce trap levels within the bandgap [8,9]. These trap states selectively capture one type of charge carrier, thereby delaying electron-hole recombination and resulting in enhanced PPC. Marshall et. al [8] observed a broad distribution of defect states in CVD diamond, with space charge decay behaviour that supports a trap-based model inferred from photodecay measurements. Koide et al. [10] further attributed the enhanced PPC to the buildup of a hole accumulation layer, resulting from the slow hole capture rate of boron acceptors and the presence of hole trap levels in the diamond epilayer. Additionally, Liao et. al [11] had shown that interface traps at the metal/diamond contact can also contribute to PPC. To minimize PPC, homoepitaxial single-crystalline diamond has been employed in photodetector designs. However, recent reports have revealed that PPC behavior is observed even in single-crystal H-diamond (type IIa) in the presence of surface conductivity (SC) [12]. Further, PPC has been observed in surface-conducting diamonds, induced by both air exposure and fluorofullerene adlayer [13]. Since SC in H-diamond is a shallow surface phenomenon, bulk defects are not considered a primary factor. Nevertheless, the precise mechanisms underlying the generation and relaxation of PPC in H-diamond remain unclear and require further detailed investigation.

This report presents a systematic investigation of PPC in H-diamond films by methodically varying the sheet carrier density through controlled ozonation. The study highlights the critical role of the 2DHG in the observed PPC behavior in HD. Temperature-dependent measurements reveal that PPC is influenced by random local potential fluctuations, which hinder the recombination of photoexcited carriers. Additionally, the PPC is found to be linked to percolative transport between localized states, arising from inhomogeneous hydrogen termination. These findings provide important insights into the origin of PPC, which are essential for advancing the development of diamond-based optoelectronic devices.

## 2. Experimental details

Microcrystalline diamond films used in this work were synthesised by hot filament chemical vapour deposition on $SiO_2$/Si substrate using ultrahigh pure (UHP) methane and hydrogen as feedstocks, as reported elsewhere [14]. Subsequent to growth, hydrogen termination on the diamond surface was carried out using UHP $H_2$ at a working pressure of ~ 40 mbar for 20 min, while maintaining the substrate temperature ~ 800 ºC. After this H treatment, the diamond films were allowed in hydrogen atmosphere to cool down to room temperature. Then, to modulate the surface conductivity, these HD films were subjected to ozone treatment for 60 and 90 s using a UV/ozone pro cleaner (UV ozone cleaner, Ossila). For clarity, the untreated HD film, and the 60 and 90 s ozonated samples are labelled as HD, OHD-60s and OHD-90s, respectively. Following these surface treatments, all samples were kept under ambient conditions for at least two days to allow the surface charge transfer to reach equilibrium.

The surface morphology and microstructure were studied by field emission scanning electron microscope (Supra 55, Carl Zeiss, Germany), while the quality of the crystal structure was investigated by micro-Raman spectrometer (In-via, Renishaw, UK). To measure the photoconductivity, diamond devices with interdigitated electrodes were fabricated using shadow masking method by thermal evaporation. These electrodes, each 500 μm wide and spaced ~ 300 μm apart, consist of a 20 nm layer of Pd followed by an 80 nm layer of Ag. Further, their Ohmic nature was verified through I-V characteristics. Photoconductivity measurements were performed using an Agilent B2902A precision source/measure unit, applying a constant bias of 0.2 V across the sample under illumination from a UV LED with a wavelength of ~ 400 nm and intensity of ~ 1 mW/cm$^2$. It is well-documented that UV and deep UV light can induce photon-stimulated desorption of adsorbate molecules from various surfaces. For instance, water desorption from a Pd (111) surface has been demonstrated using photons with wavelengths of 194 and 248 nm [15]. In a similar context, the current study observed a degradation of surface conductivity in HD films upon exposure to UV-C light (wavelengths below 280 nm). Even minor disturbances in the water adsorbate layer can significantly influence the surface conductivity of HD. Therefore, UV light with a wavelength of ~ 400 nm was selected for the experiments, only after confirming the stability of the surface conductivity. Additionally, temperature-dependent photocurrent measurements were carried out using a Linkam THMS600 stage over a temperature range of ~ 80 to 300 K.

## 3. Results

### 3.1. Structural analysis

Fig. 1a shows the FESEM surface morphology of the as grown HD film, with the inset showing its cross sectional view. A multi-faceted crystalline structure is clearly observable with the coalescence of grains in both lateral and vertical directions. The average grain size of the diamond film is ~ 2 μm with the thickness of ~ 2.1 μm. The surface morphology of the diamond film shows <110> texture associated with (100) and (111) planes [16]. Fig. 1b displays Raman spectrum of the as grown HD film taken at room temperature. The diamand Raman band at 1332 cm$^{-1}$ is dominant in the spectrum with full width and half maximum of ~ 7 cm$^{-1}$. Further, non-diamond carbon bands are not prominently observable in the spectrum. Hence, both the characteristics of narrow FWHM of diamond Raman band and absence of non-diamond Raman band assure the high structural qulaity of the diamond film.

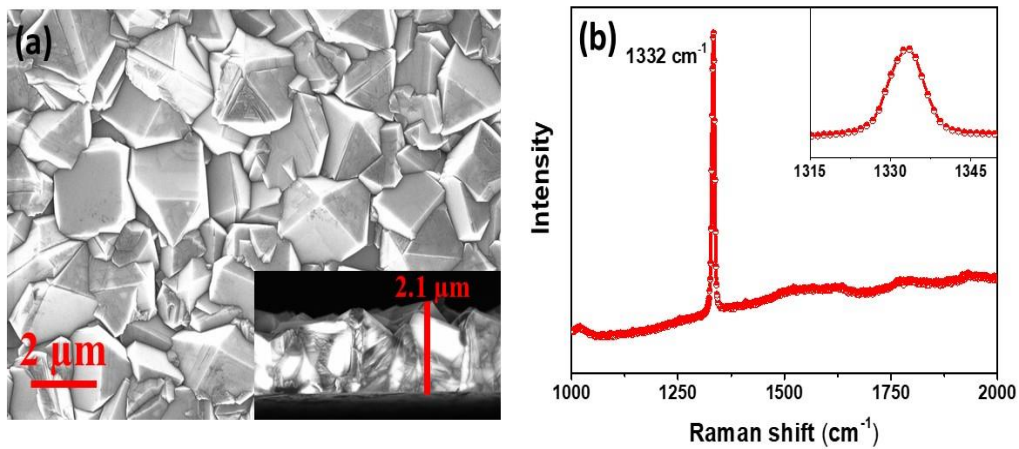

**Fig. 1.** (a) FESEM micrograph of the as-grown H-terminated diamond film. The inset displays a cross-sectional view of the film. (b) Raman spectrum of the as-grown H-diamond film, with the inset showing an enlarged view of the characteristic diamond peak near 1332 cm$^{-1}$.

### 3.2. Electrical properties

The I-V characteristics of the diamond samples HD, OHD-60s and OHD-90s confirm the Ohmic nature of the Pd/Ag metal electrodes on all three samples, as can be seen in Fig. 2a. Further, Figs. 2b & 2c clearly depict the comparison of sheet resistance and carrier density of the samples. The sheet resistance of HD, OHD-60s and OHD-90s samples are 8, 54 and 267 kΩ/□ respectively, while their carrier density is $1.1 \times 10^{13}$, $1.3 \times 10^{12}$ and $2.5 \times 10^{11}$ cm$^{-2}$ respectively. As a result of the ozonation treatment, these diamond samples exhibit nearly two orders of magnitude difference in carrier density. However, the carrier mobility of these samples does not vary significantly and lies in the range of 60 – 80 cm$^2$/Vs. The gradual increase in sheet resistance observed across the three diamond samples, following the order HD < OHD-60s < OHD-90s, is attributed to the increased presence of surface oxygen atoms, resulting from the partial substitution of surface hydrogen atoms through ozonation. On a fully hydrogen-terminated diamond surface, a dipole layer formed by C–H bonds induces a negative electron affinity (NEA) of ~ 1.3 eV. This NEA enables spontaneous electron transfer from the diamond valence band (VB) to the adsorbed water layer or other electron-accepting molecules with chemical potentials below the VB of diamond, leading to the formation of a 2DHG at the surface and sub-surface regions.

When H atoms on the diamond surface are partially placed by O atoms, local changes on both chemical and electronic structure are taking place. This leads to positive electron affinity which oppose this transfer doping. Thus, by increasing the surface coverage of O atoms on the HD, the SC and sheet carrier density can be reduced. Moreover, XPS analysis confirms that oxygen-containing functional groups on the surface of the functionalized diamond increase with ozonation [16]. Hence, by optimizing the ozonation duration for desirable H and O surface coverage, diamond samples with two orders of magnitude in variation of carrier density are achieved as discussed in the ref [16].

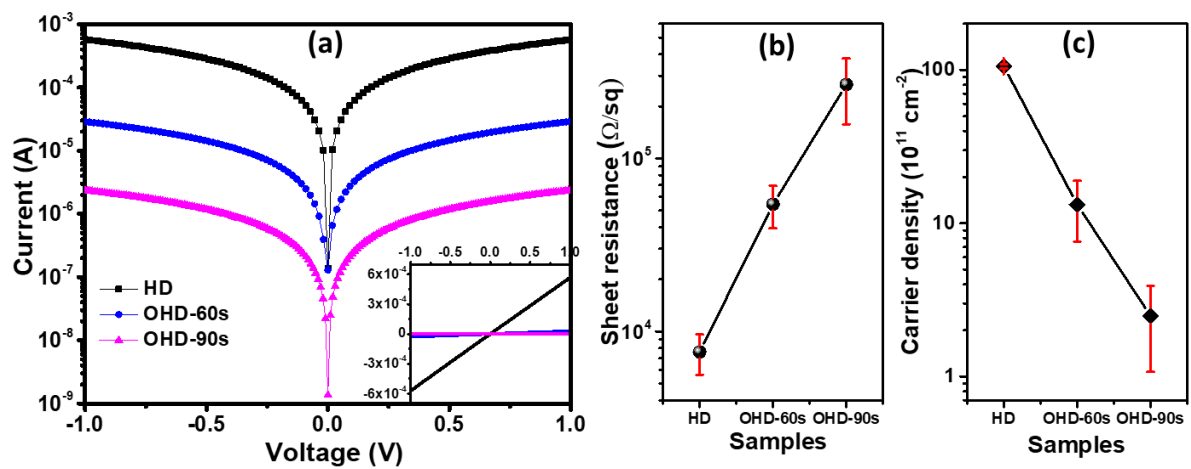

Fig. 2. (a) I-V characteristics, (b) sheet resistance and (c) carrier density of HD, OHD-60s and OHD-90s samples. The solid line in Figs. 2b and 2c is only guide to eye.

### 3.3. Persistent photoconductivity

Fig. 3 shows the photoconductivity (PC) response of diamond samples - HD, OHD-60s and OHD-90s – measured at room temperature under a sub-bandgap excitation with a wavelength of 400 nm and an applied bias of 0.2 V. The observed photocurrent in the HD samples confirms the rise in the surface hole density in the 2DHG. Note that the H-termination of diamond surface creates unoccupied surface states (SS) with energy minima positioned below conduction band (CB) minimum and also, in the mid-gap region within the bandgap [17–19]. These SS trap the excited electrons that would otherwise escape the material. Moreover, CVD-grown diamond films are also known to contain unintentional nitrogen-related defect states, which introduce multiple energy levels within the forbidden bandgap, typically ranging from 1.7 to 3.2 eV [20]. Thus, the SS related to surface H-termination and other defect states associated with bulk defects can facilitate the sub-bandgap photoconductivity by contributing the intermediate energy levels for electrons to be excited. When light with photon energy below the bandgap of diamond is exposed on its surface, electron in the VB excited to these intermediate energy level and thus, the rise in the hole density could lead to increase SC of HD surface.

As shown in Fig. 3, the photocurrent ranges from 1 to 300 µA across the samples, depending on their respective sheet resistances. For the HD sample, the photocurrent growth saturates in less than 100 seconds upon illumination, but it takes over 2800 seconds to return to its dark current level after the light is turned off. In comparison, the OHD-60s sample reaches saturation in ~ 60 seconds and requires about 1800 seconds to decay back to its dark state. The OHD-90s sample shows the fastest response, achieving saturation in under 20 seconds and recovering to its dark current level in less than 300 seconds. Overall, all three samples exhibit a rapid rise in photocurrent that reaches saturation within a few minutes of light exposure. However, when the light is turned off, the photocurrent in these samples decays in distinctly different ways. The asymmetry between the rise and decay times indicates that the samples do not follow a conventional photoconductivity process. Instead, the observed photoconductivity behavior is characteristic of PPC. Furthermore, the photoconductivity measurements were repeated three times, and the results were consistently reproducible.

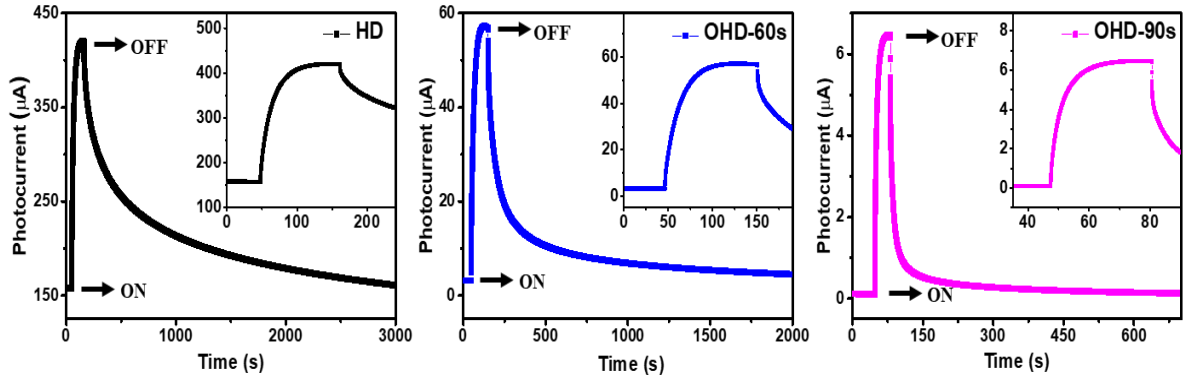

Fig. 3. The rise and decay of photocurrent as a function of time for the samples HD, OHD-60s and OHD-90s under a DC bias of 0.2 V.

To estimate the PC time constants, the photocurrent growth and decay curves are normalized and displayed in Figs. 4a and 4b respectively, for HD, OHD-60s and OHD-90s samples. The photocurrent growth curves for all three samples were best fitted using a double exponential equation of the following form.

$$I_n(t) = A_0 - B_1 e^{-\frac{t-t_0}{\tau_1}} - B_2 e^{-\frac{t-t_0}{\tau_2}}, \quad \ldots\ldots\ldots\ldots (1)$$

where, $I_n$ is photocurrent, $A_0$, $B_1$, and $B_2$ are constants, $\tau_1$, and $\tau_2$ are the time constants and $t_0$ denotes the time at which the light source was switched ON. The best fit parameters obtained from fitting the experimental data with Eq. (1) are found to be $\tau_1$= (4, 3, 1) s and $\tau_2$= (19, 14, 4) s for HD, OHD-60s, OHD-90s samples, respectively. This double exponential fitting represents that two distinct and dominant processes are involved in this photocurrent growth. Additional experimental investigations are required to identify the origin of each distinct growth process. Further, their time constants are decreasing in the order of HD > OHD-60s > OHD-90s samples and they suggest a correlation between the time constant and the sheet carrier density of the HD and partially ozonated HD films.

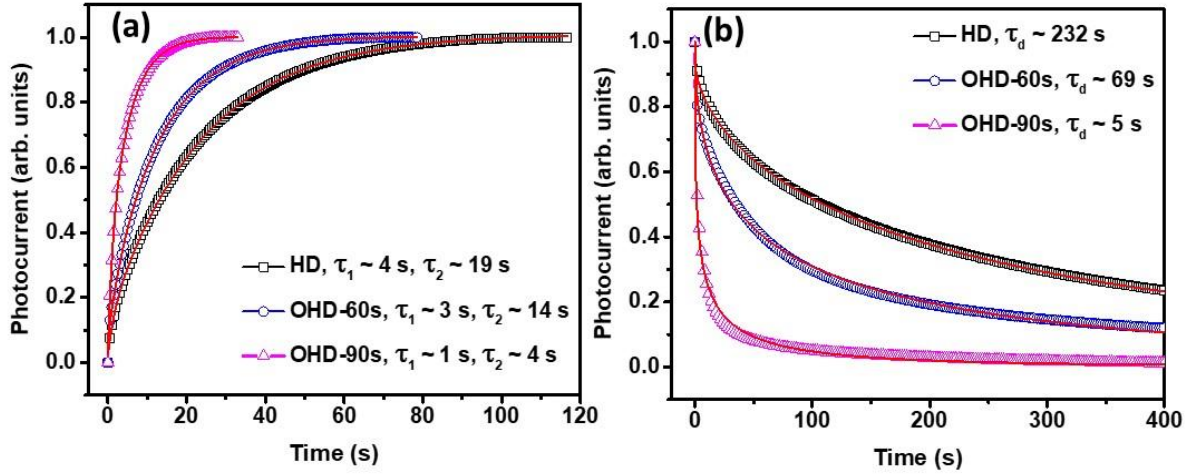

Fig. 4. The normalized photocurrent (a) growth curves and (b) decay curves for HD, OHD-60s and OHD-90s samples at room temperature. The solid red line represents the fit to the experimental data.

The photocurrent decay curves show non-exponential nature. Further, the best fit function to the decay curves is found to be stretched exponential function which is convenient to fit the decay process from single-exponential behaviour due to the combination of multiple energy transfer mechanisms [21].

$$\Delta I(t) = \Delta I_0 \exp[-(\frac{t}{\tau_d})^\beta] \quad \ldots\ldots\ldots (2)$$

where $\Delta I(t) = I(t) - I_d$, $\Delta I_0 = I(0) - I_d$, $I(t)$ is the measured current at time t, $I_d$ is the dark current, $I(0)$ is the saturated photocurrent at the moment source light is switched OFF, $\tau_d$ is the decay time and β is the stretching exponent (0 < β < 1).

As can be seen in Fig. 4b, the photocurrent decay curves of HD, OHD-60s and OHD-90s were fitted with stretched exponential function, confirming a band of charge trap states involved in the decay process. The solid red line represents the fit to the experimental data with Eq. (2). The stretched exponent β evaluated for HD, OHD-60s and OHD-90s is found to be 0.54, 0.41 and 0.38, respectively. The evaluated decay time constant $\tau_d$ is 232, 69 and 5 s for HD, OHD-60s and OHD-90s, respectively. Here, the $\tau_d$ is about 45 and 14 times higher for HD and OHD-60s samples respectively, as compared to the sample OHD-90s. Upon comparison, the $\tau_d$ is decreasing in the order HD > OHD-60s > OHD-90s. This observation also confirms the dependency of PPC effect on the sheet carrier density of the surface functionalized diamond films.

**3.4. Temperature dependent persistent photoconductivity**

The temperature dependent photocurrent decay curves of HD, OHD-60s and OHD-90s samples measured from 80 to 300 K with an applied bias of 0.2 V are shown in the Fig 5 (a, c, e). By fitting them using Eq. (2), the decay time constant $\tau_d$ and stretched exponent β at different temperature were calculated and found to decrease with decreasing temperature as shown in Fig. 5a, 5b and 5c respectively for HD, OHD-60s and OHD-90s. Further, the decay time $\tau_d$ for HD did not change significantly at lower temperatures (< 150 K) while it increases exponentially at higher temperatures. This behaviour indicates that the thermal activation mechanism is significantly reduced at low temperatures, resulting in minimal changes in $\tau_d$ compared to its behaviour at higher temperatures. Similarly, the change in the decay time $\tau_d$ for OHD-60s is insignificant for temperatures < 100 K. However, the decay time $\tau_d$ for OHD-90s continues to decrease down to 80 K, suggesting that the saturation of $\tau_d$ may occur at a temperature lower than 80 K. Hence, it is concluded that the onset temperature for increasing $\tau_d$, as a function of temperature, decreases with decreasing the sheet carrier density on the surface functionalized diamond surfaces.

Since the photocurrent decay time $\tau_d$ exhibits thermally activated behaviour above critical temperature $T_C$ in all the three samples, it is fitted with an Arrhenius plot, as shown in inset of Fig. 5 (a), (c), and (e). Here, the carrier recombination potential barrier is estimated by fitting $\tau_d$ in the temperature range of 150 – 300 K using the following Eq. 3 [22],

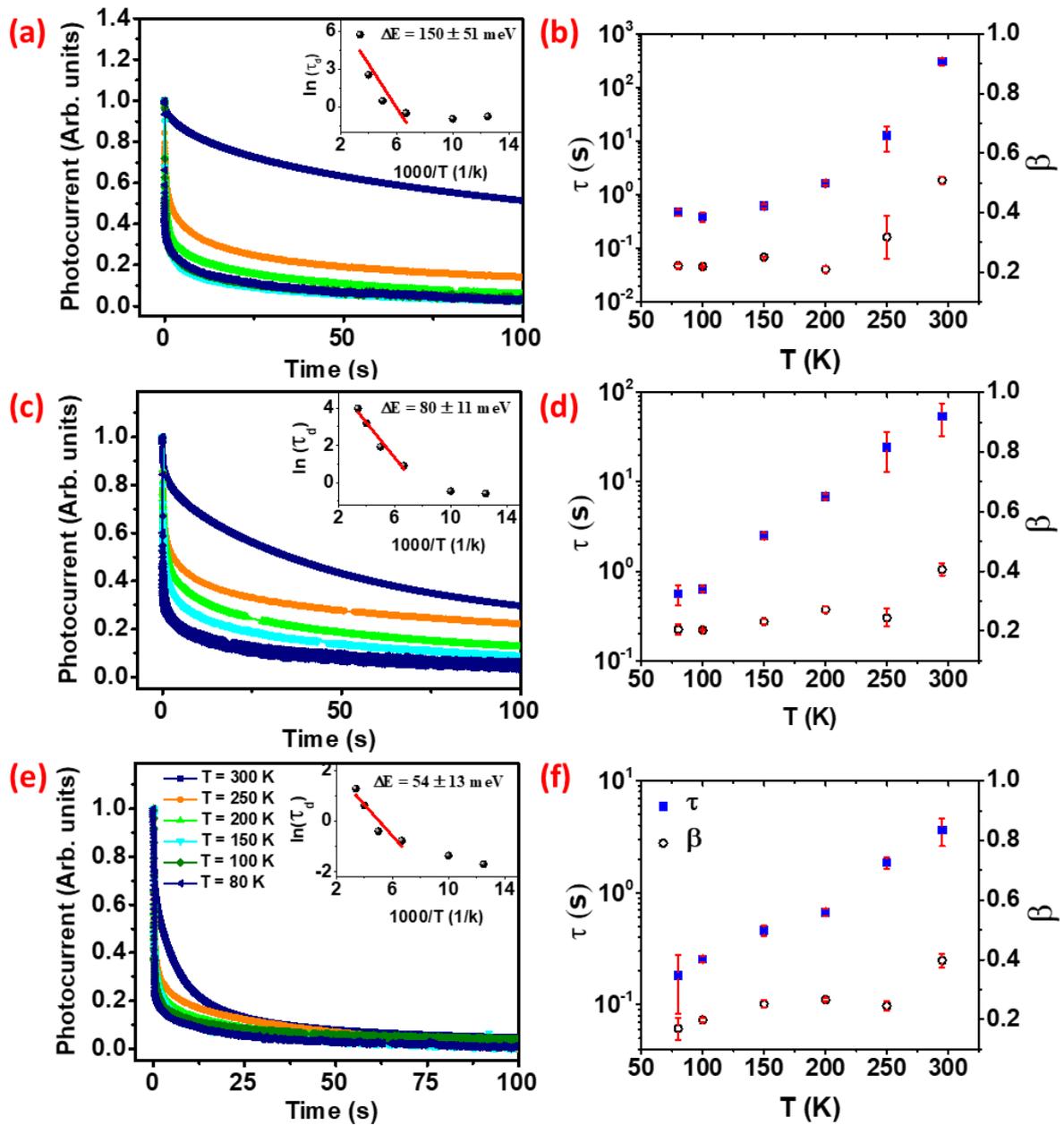

Fig. 5. (a, c, e) The normalized photocurrent decay curves at different temperatures and (b, d, f) the temperature dependence of decay time constant, $\tau_d$, and exponent, $\beta$, of HD, OHD-60s and OHD-90s samples. The insets in (a), (c) and (e) show the Arrhenius plots of $\tau_d$, which were used to calculate the activation energies.

$$\tau_d = \tau_0 \exp(\Delta E / k_B T) \quad \ldots\ldots\ldots\ldots\ldots\ldots (3)$$

where $\tau_d$ is the decay time, $\tau_0$ is a constant, $k_B$ is the Boltzmann constant, ΔE is the activation energy or carrier recombination potential barrier and T is the temperature. The carrier recombination potential barrier is evaluated to be 150 ± 51, 80 ± 11 and 54 ± 13 meV for HD, OHD-60s and OHD-90s, respectively. Upon comparison, the recombination potential barrier decreases in the order HD > OHD-60s > OHD-90s, mirroring the trend observed in sheet carrier density.

Furthermore, the temperature-dependent photocurrent is analyzed to gain insights into the conduction mechanisms in these surface-conducting diamond films. The results indicate that photocurrent transport follows a percolation model. In this regime, the temperature dependence of the PPC is expected to follow the expression given below [23].

$$I_{build-up} \propto (T - T_C)^\mu \quad \ldots\ldots\ldots\ldots\ldots\ldots (4)$$

where $I_{build\text{-}up}$ is defined as the photocurrent ($I_{light} - I_{dark}$), μ is the characteristic exponent, T is the actual temperature and $T_C$ is the critical temperature. Fig. 6 shows the temperature-dependent $I_{build\text{-}up}$ for the HD and OHD-60s samples over the range of 80 – 300 K. The solid red lines represent fitted curves based on the percolation transport model using Eq. (4). This model fits the experimental data reasonably well, and the PPC relaxation behaviour in the HD sample aligns with the predictions of the percolation model. The extracted values of the percolation exponent μ are 1.54 ± 0.04 for HD and 2.6 ± 0.49 for OHD-60s. The corresponding characteristic temperatures $T_C$_are 172 ± 1 K and 103±8 K, respectively. As discussed in Section 4.3.2, the $T_C$ for the OHD-90s sample falls below 80 K. Therefore, fitting the percolation model to the OHD-90s data within the 80–300 K range may lead to significant deviations. Nonetheless, the estimated $T_C$ values for both HD and OHD-60s samples are consistent with the transition temperatures observed in the temperature-dependent $\tau_d$ curves (Fig. 5).

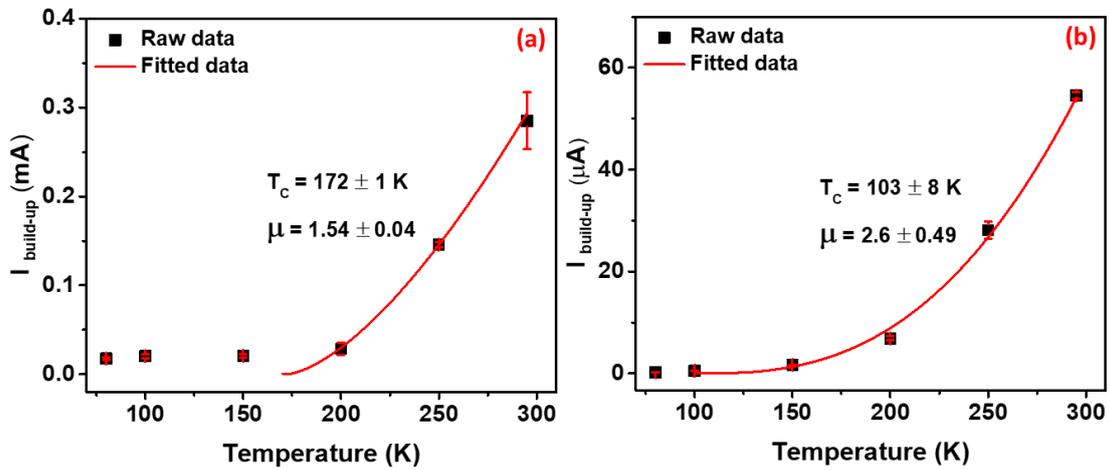

Fig. 6. The $I_{\text{build-up}}$ as a function of temperature for (a) HD and (b) OHD-60s samples. The solid red line in the plots represents the fitted curve based on percolation current transport.

## 4. Discussion

In this study, the photocurrent measurements reveal that PPC is most prominent in pristine H-diamond films and gradually decreases with the partial removal of hydrogen atoms from the surface via the ozonation process. The photocurrent decay curves for HD, OHD-60s, and OHD-90s follow a stretched exponential function, indicating the involvement of multiple energy transfer mechanisms. The decay time constant $\tau_d$ is found to be 232, 69, and 5 s for HD, OHD-60s, and OHD-90s, respectively. Moreover, $\tau_d$ exhibits thermally activated behavior above a certain critical temperature ($T_C$). Thus, an Arrhenius fitting yields carrier recombination potential barriers of ~ 150, 80, and 54 meV for HD, OHD-60s, and OHD-90s, respectively. The estimated critical temperatures, $T_C$ are ~ 172 K for the HD film and 103 K for the OHD-60s film.

Earlier, Nebel et. al [9] have attributed the origin of PPC effect in the undoped polycrystalline CVD diamond, without surface H-termination, to the bulk trap states, namely N substitutional defects, A-aggregate defects (cluster of N defects) and the defects states from grain boundaries. In their model, the trap level of 1.4 eV is estimated from the thermally stimulated currents which has a maximum around 550 K. Here, they proposed that the electrons are captured at the defect levels that are induced by grain boundaries due to $sp^2$-rich amorphous carbon and holes are mostly trapped in the A aggregated defects within the grains of the CVD diamond. Also, Liao et. al [11] have attributed the origin of PPC effect in the boron-doped homoepitaxial diamond film to the boron-induced deep defect with an activation energy of 1.37 eV with large lattice relaxation of defects. The PPC effects discussed by Nebel et. al [9] and Liao et. al [11] follow the thermally activated recombination process which contradict with the observed results here. Hence, the traps related to grain boundaries, boron defects and A aggregates are not significant for PPC observed in this work. Further, the work of Zakaria [12] had reported the PPC in the H-terminated intrinsic single crystal (100) diamonds (IIa) with the trap state photo-excitation within 2.4 eV of the VB maximum. An increase in $\tau$ with temperature, consistent with the present observations, has been previously reported above 250 K in H-terminated diamond [12]. In contrast, no such persistent photoconductivity (PPC) has been reported in the literature for O-terminated diamond. Furthermore, since the grain boundaries and other bulk defects are similar across all three films, their contribution to the observed PPC is likely minimal.

The PPC has commonly been observed across a broad range of materials, including various semiconductors and metal oxides. Several models have been proposed to explain the origin of PPC, with the most prominent being the large lattice relaxation (LLR) model, the macroscopic barrier (MB) model, and the random local potential fluctuations (RLPF) model. In the LLR model, PPC originates from deep level traps. Upon illumination, these traps transition into metastable shallow donor states, leading to the formation of a potential barrier due to the difference in lattice relaxation between the two states [24]. This barrier inhibits the recapture of photoexcited electrons, thereby sustaining the PPC. Since electron recapture is a thermally activated process, PPC in the LLR model is more pronounced at low temperatures—contrary to the behaviour observed in the studied samples. In contrast, the MB model attributes PPC to a macroscopic potential barrier that spatially separates photoexcited electron-hole pairs [7,25]. This model is typically observed in heterostructures such as quantum dot/graphene and graphene/$MoS_2$ systems [26,27]. However, the PPC described by the MB model follows a single exponential decay, which does not align with the stretched exponential decay observed in the studied samples. Therefore, our experimental data indicate that neither the LLR nor the MB mechanisms adequately explain the observed PPC in the HD samples.

In the RLPF model, the PPC is based on a decrease in the recombination rate of excited photocarriers since the carriers are separated by possible potential fluctuation. Here, the PPC is caused by the local potential fluctuations that arise due to intrinsic and/or extrinsic sources. The RLPF creates the minimum energy states near the mobility edges. The low energy carriers are localized in these spatially separated minimum energy states. At low temperatures, these carriers are well confined within the trap states, consequently, negligible PPC effect is detected. As temperature increases above a critical temperature ($T_C$), the carrier hopping between localized sites induce low-level PPC. As temperature further increases ($T > T_C$), the carriers percolate through the accessible sites with the transition of localized to delocalized state, results in the PPC effect. Hence, $T_C$ is the critical temperature at which the carriers undergo a phase transition from hopping to percolation conduction state. It is worth noting that all three surface-functionalized HD samples exhibit a nearly constant and short $\tau_d$ at low temperatures, which increases exponentially at higher temperatures—an indicative signature of the random local potential fluctuations (RLPF) mechanism [7]. Thus, the observed PPC mechanism align more closely with the RLPF model discussed here.

To understand the mechanism behind PPC in H-diamond, it is essential to examine the surface electronic structure of H-functionalized diamond. Both experimental and theoretical studies have explored the surface electronic structure of H-terminated diamond. Sque et al. [17], using ab initio density functional theory (DFT), demonstrated that the band structure of H-terminated diamond surfaces—specifically (001)-(2×1):H and (111)-(1×1):H—features a collection of unoccupied SS located below the bulk CB near the $\Gamma$ point. The minimum of the lowest unoccupied surface state lies ~ 2 eV below the CB minimum, while a second unoccupied surface state is situated near the vacuum level. Similarly, Sobaszek et al. [19] used DFT calculations to reveal a broad energy distribution of localized SS within the bandgap near the surface for both undoped and boron-doped H-terminated diamond. Chemin et al. [18], employing various spectroscopic techniques, reported that SS play a dominant role in sub-bandgap charge transfer in H-terminated diamond films. Additionally, Hayashi et al. [28] identified H-related gap states in the sub-surface region of H-terminated diamond using cathodoluminescence spectroscopy. Collectively, these findings confirm that H-termination introduces SS with a broad energy distribution within the bandgap, which strongly influence the electrical and optical properties of surface-functionalized diamond.

Based on the discussion, the plausible energy band diagram of the H-terminated diamond film associated with the observed PPC mechanism—explained using the RLPF model—is illustrated in Fig. 7. The Fig 7a shows a typical energy band diagram of H-diamond upon exposure to atmospheric electron acceptor molecules such as $H_2O$ and $CO_2$. Due to the negative electron affinity of H-terminated diamond, these adsorbed molecules extract electrons from the VB, leading to the accumulation of a 2DHG near the surface and sub-surface regions.

As a result, the surface becomes p-type with high conductivity. In addition to the surface band bending, H-termination introduces a density of SS near the VB maximum, CB minimum, and within the bandgap, as indicated in Fig. 7a. Furthermore, the local density of states (LDOS) associated with these SS can vary depending on the crystallographic orientation of the H-terminated diamond surface [19].

Fig. 7b illustrates the density of states near the surface, highlighting representative energy levels including SS near the VBM, midgap states due to H-termination, nitrogen-related donor states linked to various types of point defects along with energy level of VB, vacuum and CB. Upon exposure to sub-bandgap light (~ 3.1 eV), electron–hole pairs are generated through various possible electronic transitions, as shown in Fig. 7b. These transitions include, (a) from the VB to midgap SS or nitrogen-related donor levels ($E_N$) or vacuum level, (b) from the SS near Fermi level to midgap SS or $E_N$ levels or vacuum, and the transition to CB from (c) vacuum, (d) midgap SS and (e) $E_N$. These transitions are indicated by dotted double-arrow lines labeled a, b, c, d, and e in Fig. 7b. However, the generated minority carriers may be trapped by SS or $E_N$ levels before they can recombine, contributing to the PPC effect. As we have already ruled out the traps related to $E_N$ states, the observed PPC must be related SS associated with H-termination.

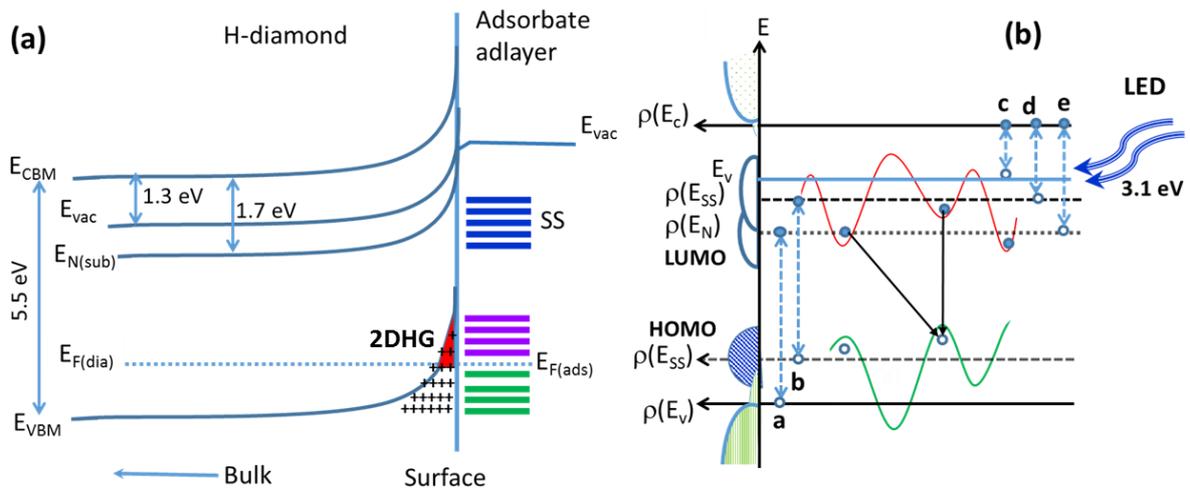

Fig. 7. (a) Schematic energy band diagram of H-terminated diamond upon exposure to adsorbate molecules, illustrating the modulation of surface band bending, 2DHG accumulation, and surface states. (b) The density of states near the surface and possible intra-band electronic transitions induced by a sub-bandgap light source. The variations in the energy band near the Fermi level and midgap states, induced by random potential fluctuations, are also illustrated.

Since the PPC behavior in HD films follows the RLPF model, it implies the presence of random local potential fluctuations characterized by multiple energy maxima and minima within the valence band and midgap states, as shown in Fig. 7b. The possible origins of these fluctuations in HD films can be attributed to inhomogeneous charge distribution, arising from intrinsic factors such as inhomogeneous hydrogen termination and/or extrinsic factors like non-uniform surface adsorbates. In hydrogenated diamond, surface adsorbates play a crucial role in controlling surface conductivity unlike in many conventional semiconductors. Since adsorbate distribution is highly sensitive to surface termination, even small variations in H-termination can have a significant impact. Here, the hydrophobic surface of H-diamond can bind more water molecules but they interact weakly and exhibit diffusive behavior due to the lack of strong bonding [29,30]. Under such conditions, slight differences in hydrogen coverage can lead to uneven adsorbate distribution, further enhancing spatial charge inhomogeneity. These observations strongly indicate that inhomogeneous hydrogen termination is the primary contributor to the potential fluctuations observed in the films.

The primary factors contributing to inhomogeneous hydrogen termination on the diamond surface include incomplete H-termination, C–H dipole disorder, and structural surface inhomogeneity [31]. Achieving 100% H-termination is inherently challenging due to steric repulsion between adjacent hydrogen atoms on the surface [32]. In addition, surface roughness and the presence of grain boundaries introduce significant structural inhomogeneities across the diamond surface [33]. These imperfections—such as non-uniform termination and surface defects—can result in various bonding configurations, including tangled structures, $-CH_2-$ and $-CH_3$ groups, and localized surface reconstructions. Collectively, these forms of disorder generate localized hole states at the diamond surface [24,27]. Due to variations in both surface termination and surface adsorbates, different regions of the surface hold different amounts of charge. This leads to spatially varied band bending with the formation of local potential fluctuations, which ultimately leads to the manifestation of PPC in H-terminated diamond films.

According to the RLPF model [23,34], both the $T_C$ and the $E_{rec}$ are determined by the extent of local potential fluctuations within the system. In the studied HD films, both $T_C$ and the $E_{rec}$ are observed to increase with rising surface carrier density. This correlation aligns with the understanding that the amplitude of local potential fluctuations intensifies as the carrier density at the surface increases. These fluctuations are further influenced by Coulombic interactions between the positively charged 2DHG and the negatively charged adsorbate layer. As the inhomogeneity associated with adsorbates with acceptor charges increases, these interactions become stronger, deepening the potential fluctuations and enhancing carrier confinement [35,36]. Consequently, the increased amplitude of these fluctuations leads to higher recombination barriers and elevated critical temperatures, as observed in HD samples. Furthermore, partial substitution of H- with O- on H-diamond surfaces results in a shift of the local electron affinity toward positive values, accompanied by an increase in surface work function and the formation of localized defect or trap states. With increasing oxygen concentration, the surface exhibits enhanced nanoscale inhomogeneity in terms of local electron affinity, surface potential, and oxygen-related electronic states, as evidenced by Kelvin probe force microscopy and other surface characterization techniques [37, 38]. Additionally, the formation of C–O bonds helps reduce inhomogeneous adsorbates carrying acceptor-like charges, while simultaneously increasing the density of surface states associated with partial oxygen termination. These oxygen-related surface states serve as efficient recombination centres, thereby accelerating photoconductivity decay. Consequently, the magnitude of persistent photocurrent diminishes with increasing levels of surface oxidation. Moreover, the positions of energy maxima and minima within the bandgap shift with temperature. Below $T_C$, charge carriers (electrons or holes) remain tightly bound within these potential wells, and conduction occurs via hopping between localized states. Under such conditions, the PPC effect is negligible. However, above $T_C$, the system transitions to a regime where conduction follows a percolation mechanism, as illustrated in Fig. 6. Therefore, by modulating the surface adsorbate molecules via partial O-termination, it is possible to tune the magnitude of the PPC in functionalized diamond surfaces.

## 5. Conclusions

In summary, we have demonstrated that persistent photoconductivity (PPC) in surface-conducting hydrogenated diamond (HD) can be effectively tuned through oxygen termination, which alters the surface adsorbates and, consequently, the surface electronic structure. The photocurrent decay curves exhibit a non-exponential behavior and are well described by a stretched exponential function. Additionally, the temperature-dependent PPC behavior in HD aligns with the random local potential fluctuation model, where fluctuations arise from inhomogeneous surface charge distribution. The estimated recombination barriers are 150, 80, and 54 meV for the pristine, 60 s, and 90s ozonated HD samples, respectively—showing a clear decrease with increasing surface oxygen concentration. Furthermore, the PPC is closely linked to percolation transport between localized states created by the inhomogeneous H-termination. According to the percolation model, the critical temperature $T_C$, which marks the transition from hopping to percolative conduction, also decreases with increasing surface oxygen concentration. The dependence of PPC on partial O-termination in HD is attributed to Coulombic interactions between the two-dimensional hole gas (2DHG) and the adsorbate layer. Overall, this study offers valuable insights into the PPC mechanism in HD, which are crucial for advancing the development of HD-based optoelectronic devices.